\documentstyle[epsf]{mn}
\newif\ifAMStwofonts
\ifoldfss
  \ifCUPmtlplainloaded \else
    \NewTextAlphabet{textbfit} {cmbxti10} {}
    \NewTextAlphabet{textbfss} {cmssbx10} {}
    \NewMathAlphabet{mathbfit} {cmbxti10} {} % for math mode
    \NewMathAlphabet{mathbfss} {cmssbx10} {} %  "   "    "
  \fi
  \ifAMStwofonts
    \ifCUPmtlplainloaded \else
      \NewSymbolFont{upmath} {eurm10}
      \NewSymbolFont{AMSa} {msam10}
      \NewMathSymbol{\upi}     {0}{upmath}{19}
      \NewMathSymbol{\umu}     {0}{upmath}{16}
      \NewMathSymbol{\upartial}{0}{upmath}{40}
      \NewMathSymbol{\leqslant}{3}{AMSa}{36}
      \NewMathSymbol{\geqslant}{3}{AMSa}{3E}

       \let\le=\leqslant
       
    \fi
  \fi
\fi % End of OFSS

\ifnfssone
  \newmathalphabet{\mathit}
  \addtoversion{normal}{\mathit}{cmr}{m}{it}
  \addtoversion{bold}{\mathit}{cmr}{bx}{it}
  \newmathalphabet{\mathbfit} % math mode version of \textbfit{..}
  \addtoversion{normal}{\mathbfit}{cmr}{bx}{it}
  \addtoversion{bold}{\mathbfit}{cmr}{bx}{it}
  \newmathalphabet{\mathbfss} % math mode version of \textbfss{..}
  \addtoversion{normal}{\mathbfss}{cmss}{bx}{n}
  \addtoversion{bold}{\mathbfss}{cmss}{bx}{n}
  \ifAMStwofonts
    \ifCUPmtlplainloaded \else
      \UseAMStwoboldmath
      \makeatletter
      \new@mathgroup\upmath@group
      \define@mathgroup\mv@normal\upmath@group{eur}{m}{n}
      \define@mathgroup\mv@bold\upmath@group{eur}{b}{n}
      \edef\UPM{\hexnumber\upmath@group}
      \new@mathgroup\amsa@group
      \define@mathgroup\mv@normal\amsa@group{msa}{m}{n}
      \define@mathgroup\mv@bold\amsa@group{msa}{m}{n}
      \edef\AMSa{\hexnumber\amsa@group}
      \makeatother
      \mathchardef\upi="0\UPM19
      \mathchardef\umu="0\UPM16
      \mathchardef\upartial="0\UPM40
      \mathchardef\leqslant="3\AMSa36
      \mathchardef\geqslant="3\AMSa3E

       \let\le=\leqslant

    \fi
  \fi
\fi % End of NFSS release 1

\ifnfsstwo
  \DeclareMathAlphabet{\mathbfit}{OT1}{cmr}{bx}{it}
  \SetMathAlphabet\mathbfit{bold}{OT1}{cmr}{bx}{it}
  \DeclareMathAlphabet{\mathbfss}{OT1}{cmss}{bx}{n}
  \SetMathAlphabet\mathbfss{bold}{OT1}{cmss}{bx}{n}
  \ifAMStwofonts
    \ifCUPmtlplainloaded \else
      \DeclareSymbolFont{UPM}{U}{eur}{m}{n}
      \SetSymbolFont{UPM}{bold}{U}{eur}{b}{n}
      \DeclareSymbolFont{AMSa}{U}{msa}{m}{n}
      \DeclareMathSymbol{\upi}{0}{UPM}{"19}
      \DeclareMathSymbol{\umu}{0}{UPM}{"16}
      \DeclareMathSymbol{\upartial}{0}{UPM}{"40}
      \DeclareMathSymbol{\leqslant}{3}{AMSa}{"36}
      \DeclareMathSymbol{\geqslant}{3}{AMSa}{"3E}

       \let\le=\leqslant

    \fi
  \fi
\fi % End of NFSS release 2

\ifCUPmtlplainloaded \else
  \ifAMStwofonts \else % If no AMS fonts
    \def\upi{\pi}
    \def\umu{\mu}
    \def\upartial{\partial}
  \fi
\fi

\title[Lithium in Host Stars of Extrasolar Planets]{The Host Stars of Extrasolar
Planets Have Normal Lithium Abundances}
\author[Sean G. Ryan]{Sean G. Ryan\\
Dept of Physics and Astronomy, The Open University, Walton Hall, 
Milton Keynes MK7 6AA, UK.
email: s.g.ryan@open.ac.uk}
\date{Accepted YEAR MONTH DATE.
      Received YEAR MONTH DATE;
      in original form YEAR MONTH DATE}

\pagerange{\pageref{firstpage}--\pageref{lastpage}}
\pubyear{000614}
\begin{document}

\maketitle

\label{firstpage}

\begin{abstract}
The lithium abundances of planet-harbouring stars have been compared  with those
of open clusters and field stars. 
Young (chromospherically active) and subgiant stars have been eliminated from
the comparison because they are at different stages of evolution
and Li processing to the planet-harbouring stars, and hence have systematically
higher Li abundances.
The analysis showed that the Li abundances of the planet-harbouring stars
are indistinguishable from those of non-planet-harbouring stars of the
same age, temperature, and composition.
This conclusion is opposite to that arrived at by Gonzalez \& Laws (2000);
it is believed that the field star sample used by them contained too wide
a range of ages, evolutionary types, and temperatures to be accommodated by the
model they
adopted to describe the dependence on parameters. 
Li does not appear set to provide key insights into the formation and evolution
of planetary systems.
\end{abstract}


\section{Introduction}					% 1

The discovery of tens of extra-solar planets over recent years
(Wolszczan 1994; Mayor \& Queloz 1995; Marcy \& Butler 2000)
has re-invigorated efforts to understand the processes by which planetary systems form.
The existence of more than one system --- the solar system --- to study and
the prospect of many more being discovered has spurred
this effort.  

One way to help understand the formation of planetary systems is to discover
characteristics which distinguish planet-harbouring stars from
lone stars. 
They are more metal rich than the general stellar population 
(Fuhrmann, Pfeiffer \& Bernkopf 1997; Gonzalez 1997, 1999), 
and the difference between 
solar photospheric and meteoritic abundances correlates with elemental 
condensation temperature, consistent with self-enrichment of the solar surface
(Gonzalez 1997).
Gonzalez (1999) discusses the anomalously small velocity of 
the sun relative to the local standard of rest (LSR), but the explanations 
are based on anthropic arguments which do not tell 
us about other planetary systems. Genuine characteristics
not only provide information to help understand the formation of these
systems, but could also help bias future searches towards
planet-harbouring systems of that type.

Perusal of the characteristics of exoplanet hosts can give the impression that
they are unusually Li deficient compared to lone stars. 
Several stars now known to have planetary
systems were flagged as having low Li abundances {\it prior} to the
discovery of their companions. HR~5968 ($\rho$ CrB) was 
singled out by Lambert, Heath \& Edvardsson (1991), and
Friel et al. (1993) commented on the large Li difference 
between 16 Cyg A and B despite their similar
temperatures, though they did not suggest that processes other than normal
single-star evolution would be needed to explain the lower abundance
in 16 Cyg B. As a third example, the low Li abundance in the solar photosphere
($A$(Li)~=~1.10$\pm$0.10; Grevesse \& Sauval 1998) compared with the pre-solar 
nebula ($A$(Li)~=~3.31$\pm$0.04 in meteorites) has long challenged standard
stellar evolution models (e.g. Deliyannis 1995). ($A$(Li) $\equiv$ 
log$_{10}$ ($n$(Li)/$n$(H)) + 12.00.)

Lithium is special because stars
destroy it 
during pre-main sequence and main-sequence evolution, depending on their mass
and metallicity. When surface material is mixed down to depths where the
temperature exceeds 2.5$\times$10$^6$~K, Li-purged material is returned to the 
surface. 
Li survival therefore reflects the mixing history,
and in the context of planet-harbouring stars could provide 
information on the accretion of material and the angular-momentum evolution of 
the system as a whole.

\begin{table*}
 \centering
%\begin{minipage}{140mm}
 \begin{minipage}{170mm}
 \caption{Host stars to planetary systems in which lithium has been measured}
% \begin{tabular}{@{}llccrccccrccl@{}}
  \begin{tabular}{@{}lllccrccccrccl@{}}
%  {Star}& &{$T_{\rm eff}$}&{$\epsilon_T$}&{[Fe/H]}&{$\epsilon_{\rm [Fe/H]}$}&
   {Star}& & &{$T_{\rm eff}$}&{$\epsilon_T$}&{[Fe/H]}&{$\epsilon_{\rm [Fe/H]}$}&
   {age}&{$M_V$}&{$\epsilon_{M{\rm v}}$}&{$A$(Li)}&{$\epsilon_{A{\rm (Li)}}$}&
   log$R'_{\rm HK}$&{Refs}\footnote{
References:
(a)  Gonzalez \& Laws 2000;
(b)  King et al. 1997;
(c)  Boesgaard \& Lavery 1986;
(d)  Lambert et al. 1991;
(e)  Pasquini et al. 1994;
(f)  Edvardsson et al. 1993;
(g)  Duncan 1981;
(h)  Rebolo et al. 1988;
(i)  Gonzalez et al. 1999;
(j)  Fran\c cois et al. 1996;
(k)  Gonzalez 1998;
(l)  Friel et al. 1993;
(m)  Randich et al. 1999;
(n)  Gonzalez \& Vanture 1998;
(o)  Favata et al. 1997.
}
   \\
%  {    }& &{    K        }&{    K       }&{      }&{                       }&
   {    }& & &{    K        }&{    K       }&{      }&{                       }&
   {Gyr}&{ mag }&{ mag                 }&{       }&{                        }&
   { }&{                        }
   \\[10pt]
HR5185 &HD120136&$\tau$ Boo     &6420& 80&  0.32& 0.06 & 1.5&3.53&0.03&  1.68  &0.25&      &a,b,c\\
HR3947 &HD75289 &       &6140& 50&  0.28& 0.05 & 2.1&4.04&0.04&  2.76  &0.05&--5.00&a\\
HR458  &HD9826  &$\upsilon$ And &6140& 60&  0.12& 0.05 & 3.3&3.45&0.03&  2.26  &0.07&--4.97&a,d\\
HR810  & HD17051&       &6074&100&--0.04&      &    &4.22&0.02&  2.39  &    &--4.65&e\\
HR4277 &HD95128 &47 UMa         &5882& 40&  0.01&      & 6.9&4.29&0.02&$<$1.70 &    &--4.95&b,f,g\\
HD114762&       &        &5870& 40&--0.74& 0.03 &13.8&4.26&0.13&  1.92  &    &      &d,f,h\\
HD187123&        &       &5830& 40&  0.16& 0.05 &  4 &4.43&0.08&  1.20  &0.20&      &a i\\
HR8729 &HD217014 &51 Peg         &5777& 40&  0.06&      & 8.5&4.52&0.03&  1.16  &0.05&--4.97&b,f,j\\
Sun    &        &       &5770&   & =0.00& =0   & 4.5&4.72&    & 1.10   &0.10&      &a,b\\
HR5968 &HD143761 &$\rho$ CrB     &5750& 50&--0.35& 0.06 & 11 &4.18&0.03&  1.30  &0.10&--5.02&f,d,k\\
HD186427&16CygB       &       &5747& 20&  0.05& 0.06 &    &4.60&0.02&$<$0.60 &    &      & l,b\\
HR8734 &HD217107       &       &5597&   &  0.30&      & 10 &4.71&0.03&$<$0.64 &    &      & m\\
HD210277     &  &       &5540& 60&  0.24& 0.05 & 12 &4.90&0.05&$<$0.80 &    &      &a i\\
HR5072 &HD117176 &70 Vir         &5500&   &--0.11&      &    &3.68&0.03&  1.12  &    &--5.11&g,b\\
HD145675 &14 Her       &         &5300& 90& 0.50 & 0.05 &  6 &5.32&0.03&$<$0.70 &    &--5.10& i,a\\
HR3522 &HD75732&$\rho^1$ 55 Cnc&5250& 70&  0.45& 0.05 &    &5.47&0.02&$<$0.46 &0.15&--4.97& n,a\\
HR637 & HD13445&   GJ86        &5072&   &      &      &    &5.93&0.01&$<-$0.24&    &--4.74&  o\\
\end{tabular}
\end{minipage}
\end{table*}

Li deficiency in
planet hosts was assessed by King et al. (1997) and Gonzalez \& Laws (2000).
King et al. examined 16 Cyg A and B, and commented on 
six other systems.
% 47~UMa, 51~Peg, the sun, 
 HD~114762, 70~Vir, and $\tau$~Boo. 
They concluded that ``the data are too few at this point to 
establish a connection between alleged  planetary companions and photospheric 
Li abundances'', whilst acknowledging ``It is possible, in principle anyway, 
that the low Li abundances ... may be
related to the presence of a planetary companion.''
Gonzalez \& Laws concluded more positively that
``stars with planets tend to have smaller Li abundances when corrected for
difference in $T_{\rm eff}$, [Fe/H], and $R'_{\rm HK}$'' (where 
$R'_{\rm HK}$ is a chromospheric emission measure).

The current study was prompted by the cases of HR~5968, 16 Cyg A and B,
and the Sun, independently of the work by King et al. and Gonzalez \& Laws.
However, the opposite conclusion was reached compared to that of 
Gonzalez \& Laws.
Instead it showed that the Li abundances of planet-harbouring stars are
indistinguishable from those
of otherwise similar lone stars. The arguments leading to this 
negative conclusion will be presented in this paper.

\section{Data}

All data in this study were taken from the literature. Extensive
use was made of {\it Simbad} and the online {\it Hipparcos} catalog 
(ESA 1997) 
provided by the CDS.
Planet-harbouring stars\footnote{see 
http://cfa-www.harvard.edu/planets/catalog.html}
for which Li abundances have been published are listed in
Table~1.
Where Li abundances are available from more than one source, the most recent has
been adopted.
Most have $-0.35~\le$~[Fe/H]~$\le~0.45$. 
Figure~1(a) gives the HR diagram based on accurate {\it Hipparcos} parallaxes,
while Figure~1(b) shows $A$(Li) vs $T_{\rm eff}$.

\begin{figure}

% \vspace{150mm}

\epsfxsize=080mm
\epsfbox[44 170 322 690]{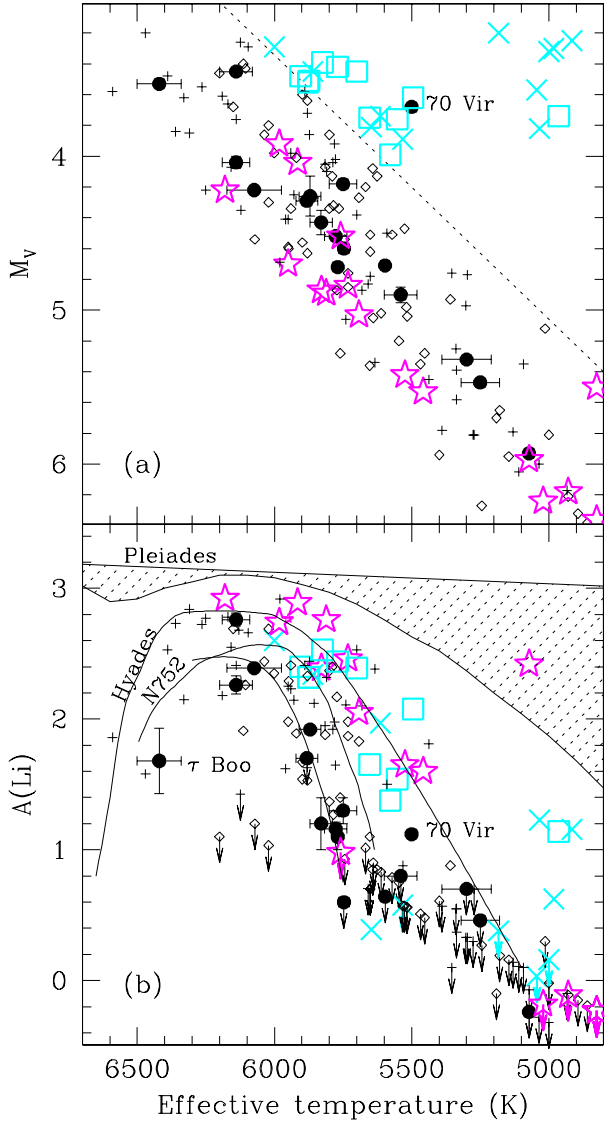}
% \epsfbox[44 170 322 690]{liplf1bw.ps}

 \caption{
(a) HR diagram for: {\it solid symbols} planet host stars;
{\it squares} and {\it crosses} field subgiants of low and unknown chromospheric
activity; 
{\it diamonds} and {\it plus-signs} main-sequence field stars of low and unknown
activity; 
and {\it five-pointed stars} known chromospherically active objects.
The dotted line separates main-sequence and subgiant stars.
(b) Li abundances for sample in (a). The shaded zone is the region
occupied by Pleiades dwarfs. The other fiducials are
(in descending order) for the Hyades, NGC 752, and M67.
In both panels, error bars are shown only for the planet hosts, to 
aid clarity. Subgiants and chromospherically-active stars, which have been 
emphasised, have higher Li abundances; see text.
}
\end{figure}

Open clusters and field stars of appropriate age, temperature and metallicity
can be used to reveal the ``normal'' evolution of Li. 
Shown in Figure~1(b) are fiducial lines (Hobbs \& Pilachowski 1988;
Ryan \& Deliyannis 1995) 
for the Pleiades, Hyades, NGC 752 and M67, whose parameters are given in 
Table~2.
Field stars whose Li 
abundances are known from Lambert et al. (1991), 
Pasquini, Liu \& Pallavicini (1994), Favata, Micela \& Sciortino  (1996, 1997), and 
Randich et al. (1999) are also shown. Two other stars have been added for 
reasons that will become clear later: 16~Cyg~A and $\alpha$~Cen~A.

\begin{table}
 \centering
 \caption{Open cluster ages and metallicities}
  \begin{tabular}{@{}lcc@{}}
   {Cluster}&{Age/Gyr}&{[Fe/H]}
   \\[10pt]
% Cluster       age/Gyr [Fe/H]^M
Pleiades        &0.08   & 0.0 \\
Hyades          &0.7    & 0.1 \\
NGC 752         &1.7    & 0.0 \\
M67             & 5     & 0.0 \\
\end{tabular}
\end{table}

Several criteria have been to restrict the field stars used in the 
comparison sample. 
Firstly, only objects with absolute magnitudes from {\it Hipparcos}, typically 
accurate to  $\pm$ 0.03--0.10~mag, have been admitted. This is so their
evolutionary states are known.
Secondly, the most luminous of the planet-hosting stars has
$M_{\rm V} = 3.45 \pm 0.03$, so field stars having $M_{\rm V} < 3.20$ were 
excluded.
Thirdly, stars lying outside the range $-0.35~\le$~[Fe/H]~$\le~0.45$, 
the same as the majority of the planet-harbouring sample, have been rejected
to reduce the impact of 
stars having formed at different stages of Galactic chemical evolution
(e.g. Ryan et al. 2000). Favata et al. (1996, 1997) do not tabulate 
metallicities; values from Cayrel de Strobel et al. (1997) 
have been used where possible.

\section{Analysis}

The open cluster fiducials show that the youngest clusters have
higher Li abundances, despite having
similar metallicities. The steepness of 
the depletion curve,  d$A$(Li)/d$T_{\rm eff}$, also depends on age. 
Table~1 shows that age estimates (where they exist) for the planet host stars
range from 1.5 -- 14~Gyr, so they should lie below the NGC 752 fiducial.
However, the ages of field stars are difficult to derive accurately. A useful 
surrogate 
for age in {\it young} Population I stars is chromospheric activity;
the youngest stars show greater activity.
The distribution of the Ca~{\small II} H and K line-core emission diagnostic,
log~$R'_{\rm HK}$, in the study by Henry et al. (1996, Fig.~8) 
is strongly bimodal. Some 70\% of the stars of {\it their} sample constitute
in an inactive peak from  $-5.50 <$~log~$R'_{\rm HK} < -4.65$, the remainder
having higher activity levels $-4.65 <$~log~$R'_{\rm HK}~^<_\sim -4.0$.
Measures of chromospheric activity from Soderblom (1985) and Henry et al.
(1996) are available for ten of the planet hosts, and all fall
within the lower activity peak, the highest level being log~$R'_{\rm HK} = -4.65$ 
(HD 17051) at the local minimum in the bimodal distribution. 
Measurements are also available for many of the non-planet-harbouring 
stars. (Pasquini et al. (1994) measure a different chromospheric 
emission measure, $F'_k$. A least squares fit to
stars in both surveys yielded the transformation
log~$R'_{\rm HK} = 0.755~F'_k - 9.141$.)

Attempting to account for variations in the Li abundance with
age, metallicity, and effective temperature, 
Gonzalez \& Laws (2000) performed a fit to a similar sample of
field stars using an equation $A$(Li) = 
$a_0$ + $a_1$[Fe/H] + $a_2$~log~$R'_{\rm HK}$ + $a_3$ log~$T_{\rm eff}$.
The approach adopted in the present work differs; a polynomial of this form is 
regarded as inappropriate. Instead, an effort is made to
eliminate stars whose parameters do not coincide with the planet-host
sample, and then to compare the stars in the $A$(Li) vs $T_{\rm eff}$ plane
directly. As will be shown below, the approach adopted here leads to the 
opposite conclusion to the one reached by Gonzalez \& Laws. 

At first glance, Figure~1(b) seems to justify the belief that planet hosts
have lower Li abundances. However, the
non-planet-harbouring sample in Figure~1(b) is {\it not} broadly similar to
that of the planet hosts. Two groups of unrepresentative stars have been
highlighted.
{\it Star} symbols indicate objects whose activity exceeds 
log~$R'_{\rm HK} = 4.65$, or $F'_k = 6.12$, 
the highest measurement for planet-hosts (HD 17051).
This coincides with the local minimum in Henry et al's bimodal 
distribution.
Figure~1(a) verifies that these are generally less luminous,
typical of young stars lying closer to the zero age main sequence. 
Figure 1(b) shows 
that their lithium abundances are amongst the highest in the sample. 
Although Gonzalez \& Laws (2000) attempted to fit this dependence, 
the approach here is instead to eliminate those stars entirely.
This reduces the chance of comparing un-alike samples. Furthermore, it is unclear
that $A$(Li) depends linearly on this parameter.
There are examples in Figure~1(b) where ``active'' stars having the same
$T_{\rm eff}$ have very different $A$(Li) values;
a linear model cannot fully capture the effect. 
Instead, here such stars are eliminated
as unrepresentative of the population of less-active planet hosts.
Note, however, that this elimination is incomplete, as there are stars for which
chromospheric diagnostics are lacking. They are shown as {\it crosses} and 
{\it plus} signs for subgiant and main-sequence stars.
It is likely that some of the latter with low luminosities and 
high Li abundances would be
eliminated if more complete data were available.

Secondly, Hipparcos parallaxes
allow us to distinguish main sequence stars from subgiants, which
owe their Li destruction to different processes
Ryan \& Deliyannis (1995). The former mix surface 
material to depths, greater in cooler stars, where it is {\it destroyed} at 
$T > 2.5\times10^6$~K.
Subgiants had higher temperatures when they were on the main sequence, and
either may have experienced less Li destruction, or at the other
extreme may have depleted Li extensively if located between
6400 and 6900~K at the F-star Li gap (Boesgaard \& Tripicco 1986). 
Once on the subgiant branch, they
{\it dilute} surface Li as deepening convection mixes Li-purged 
material up to the surface; {\it dilution} without additional {\it destruction}
occurs initially.
In Figure~1(a), stars are defined as subgiants if they fall in the region at 
upper right defined by $M_{\rm V} < 13.63 - 1.7143\times10^{-3} T_{\rm eff}$,
and are shown as {\it squares} and {\it crosses} depending on whether or not
chromospheric-activity measurements are available. (Chromospheric activity
is not expected in normal subgiants.) They are seen (Figure 1(b))
to have higher $A$(Li) values than main-sequence stars.
The two groups must be analysed separately;  there is no indication whether
Gonzalez \& Laws (2000) made this distinction.

There is only one subgiant planet-host in this study, 70~Vir. With 
$T_{\rm eff} = 5500$~K, it lies along the trend towards diminishing
$A$(Li) at lower $T_{\rm eff}$, coincidentally close to the Hyades
fiducial. The subgiants with $T_{\rm eff} < 5700$~K exhibit a wide range of
Li abundances; 70~Vir sits in the middle of that range, giving no indication 
that it is abnormally Li-poor.
The wide range of Li abundances in these subgiants may
arise because some of them lay in the wings of 
the F-star Li dip when they were on the main sequence. 	
As Gonzalez \& Laws recognised, the low Li abundance of $\tau$~Boo is certainly
due to the Li dip.

Figure 2, from which {\it known} chromospherically active stars and subgiants 
(but not 70~Vir) have been eliminated, contains main-sequence stars.
If we adopt log~$R'_{\rm HK} = -4.9$ (the modal value of 
Henry et al's activity distribution) as characteristic of inactive stars,
and compute the Li abundance from Gonzalez \& Laws' (2000, eq(1)) fit for
[Fe/H] = $-0.3$ and +0.2, the dashed lines in Figure~2 are obtained. These fail
to fit the decreasing Li abundance in the cooler main-sequence 
field or open cluster stars. This is almost certainly due to the
elimination of inappropriate objects (known chromospherically-active stars
and subgiants) in the present work. 
The fit upon which Gonzalez \& Laws based 
their conclusion was not appropriate to main-sequence, inactive stars.

\begin{figure}

% \vspace{150mm}

\epsfxsize=080mm
\epsfbox[44 170 322 450]{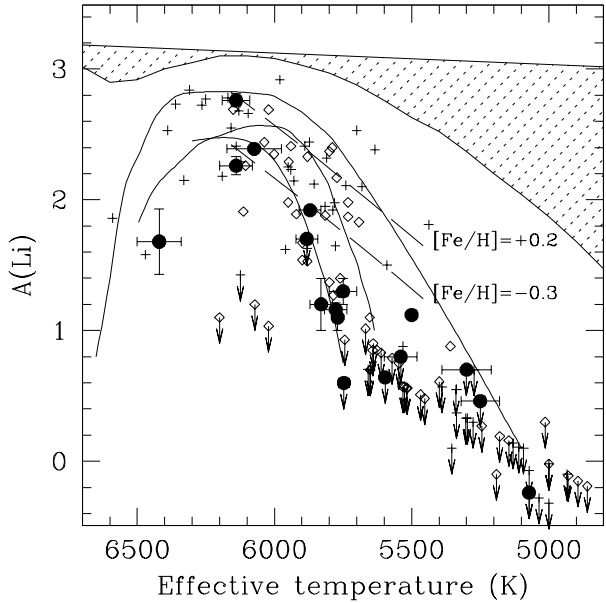}		% B&W
% \epsfbox[44 170 322 450]{liplf2bw.ps}			% = liplanet_figd.ps

 \caption{
As for Figure~1(b), but with subgiants and {\it known} chromospherically active 
stars eliminated.
Dashed lines give the model of 
Gonzalez \& Laws (2000, eq.~1) for [Fe/H] = $-0.3$ and +0.2, assuming 
chromospheric inactivity (log $R'_{\rm HK}~=~-4.9$).
The model is a poor match to this sample; see text.
  }
\end{figure}

Figure~2 can be used to reexamine whether
the Li abundances in the planet-harbouring stars are distinguishable from those
of otherwise similar stars. 
70~Vir and $\tau$~Boo have been discussed above and 
no evidence of abnormal Li deficiency found.
The three planet hosts with $T_{\rm eff} \simeq 6100$~K
are completely consistent with similar
stars in the field and the open cluster fiducials. The planet host
with the lowest Li abundance is also the most luminous, and could be an
early subgiant descended from the wing of the F-star Li dip. There is no 
evidence in these three planet hosts
for abnormally low Li abundances.

The remaining planet hosts follow the steep decline
of $A$(Li) with decreasing $T_{\rm eff}$ that both the field star samples
and the older open cluster fiducials (NGC 752 and M67) exhibit.
Whilst there exist some field stars with higher Li abundances, there also exist
many low values. Moreover, many of those with higher values have unknown
chromospheric activity levels, and it is plausible that many of these are in
fact relatively young. Considering chromospherically-inactive stars 
within $\pm$50~K of the two planet hosts at 
$T_{\rm eff} \simeq 5900$~K, one has higher $A$(Li), one has a lithium abundance
between those of the planet hosts, and three have lower Li abundances.
Widening the interval to $\pm$100~K would change the count to six above, two
between, and five below. Clearly there is nothing to distinguish these two 
planet hosts as having abnormally low Li abundances, bringing the tally
to zero Li-deficient planet hosts out of seven discussed so far.

The five planet hosts in the ``solar'' group at $T_{\rm eff} \simeq 5800$~K
provide the only hint of
possibly lower Li abundances. A count of chromospherically-inactive
stars within $\pm$50~K of the solar temperature gives eight with higher
abundances (though two only marginally and within the errorbars), two
within the $A$(Li) range of the solar group (one of which is obscured in
Figure~2), and one yielding only a
low upper limit. However, three notes of caution are required.

Firstly, the planet-harbouring stars are an excellent fit to the older open 
cluster fiducials. The youngest of these five planet hosts is HD~187123 at 
4~Gyr, and the oldest is $\rho$~CrB at 11 Gyr. All lie
close to the fiducials for NGC 752 (2 Gyr) and 
M67 (5 Gyr), so their Li abundances would be interpreted as
normal for their ages. Perhaps the high $A$(Li)
field stars within $\pm$50~K of the sun are younger and retain more Li, 
although not
so young as to remain chromospherically active.

Secondly, this $T_{\rm eff}$ is the coolest for which Li detections, as opposed 
to upper limits, are routinely measurable. The open cluster
fiducials indicate that Li depletion is a steep function of temperature,
$A$(Li) falling by 0.33~dex per 50~K. A star's ``expected'' location in the $A$(Li) vs 
$T_{\rm eff}$ plane is clearly very sensitive to the uncertainties in its 
$T_{\rm eff}$. Furthermore, the range of Li abundances even in the field sample
is 1.5~dex within this $\pm$50~K interval. It is difficult to
conclude that the planet hosts are anomalous in this circumstance.
Of particular relevance to this point is the comparison between the Sun and
$\alpha$~Cen~A 
($T_{\rm eff}$ = 5800$\pm$20, $A$(Li) = 1.37$\pm$0.06; King et al. 1997),  
and between the coeval pair 16~Cyg~A and B
($T_{\rm eff}$ = 5785 and 5747~K respectively, and 
$A$(Li) = 1.27$\pm$0.05 and $<$0.60; King et al. 1997). 
The $T_{\rm eff}$ and $A$(Li) difference between
the first two runs parallel to the NGC~752 and M67 fiducials at this 
temperature, so the difference in
$A$(Li) is entirely consistent with the different
temperatures. The rate d$A$(Li)/d$T_{\rm eff}$ for 16~Cyg~A and B 
is steeper, but they
are also marginally cooler, and as the prevalence of non-detections
(upper limits) and the steep open cluster fiducials suggest, a greater loss of 
Li in the 
coolest of these four stars would not be outrageous. 
Friel et al's (1993) emphasis on normal stellar evolutionary processes in
their comment that these two stars
``may provide a powerful constraint to models of evolution of the Li content
in solar type stars'' is simpler than postulating an
abnormal evolution of Li in stars harbouring planets.

Thirdly, if one supposes for a moment that the planet hosts
are abnormally Li-deficient, one would be struck by the great similarity in the
final abundances of the four systems 51~Peg, HD187123, $\rho$~CrB, and the Sun.
The first two planet masses and semi-major axes are
$M$~sin~$i$ $\simeq$~0.50~M$_{Jup}$ and 
$\simeq$~0.045~AU, and $\rho$~CrB has values 1.1~M$_{Jup}$ and 0.23~AU.
The parameters for the Sun are obvious.  One would be challenged to explain
why three diverse systems have similar
Li abundances if all are depleted compared to non-planet-harbouring stars. 
The alternative, that the 
four systems have the same Li abundance because that is what is natural for 
stars of their mass, age, and composition, is in accord with Occam's razor.

For the planet hosts cooler than the solar group, only upper limits on
lithium abundances are available. The same is true of almost all
field star measurements at $T_{\rm eff} < 5600$~K, so 
there is no information on the relative abundances
of planet-harbouring compared to sole stars.

\section{Conclusions}

The lithium abundances of planet-harbouring stars have been compared 
with the abundances in open clusters of known age and metallicity and with field
stars. 
Young (chromospherically active) field stars have higher Li abundances than
older stars of the same $T_{\rm eff}$, but are 
significantly younger (more active) than the planet-harbouring stars, so were 
eliminated. An examination of the $A$(Li) vs $T_{\rm eff}$ 
trends for the planet-host and field star samples were conducted separately for 
subgiants and main-sequence stars because of their different evolutionary
and Li-processing histories.
The comparisons showed no differences between the Li abundances
of the planet-host and other samples in the case of the 
planet-harbouring subgiant or the six hosts with $T_{\rm eff} > 5850$~K.
For the five solar-like planet hosts there are
examples of chromospherically-inactive lone stars having much higher Li 
abundances, but covering a huge range ($\sim$1.5~dex) in $A$(Li). 
It is likely that some of these are old enough to show no
chromospheric activity but have not yet depleted their Li abundances
to the levels seen in the older open clusters. Furthermore,
the temperature dependence of Li depletion is very high. 
The solar-temperature  planetary systems have ages greater than 
4~Gyr, and in this context their Li abundances are consistent with similarly old
open cluster and with known coeval field stars.
In particular,
the difference in $A$(Li) between $\alpha$~Cen~A and the Sun is consistent with
the decline rate d$A$(Li)/d$T_{\rm eff}$ = 0.33~dex per 50~K
inferred from 2-5~Gyr open clusters. While the 
decline rate between 16~Cyg~A and B is larger, 16~Cyg~B is cooler and
very close  to the 
temperature at which Li routinely vanishes in main-sequence stars.
In summary, there is no strong evidence that planet-harbouring
stars have lower Li abundances than open cluster stars of similar mass, age,
and metallicity, and nor are they lower than in an appropriately constituted
sample of field stars of similar age and evolutionary state. 
This conclusion is opposite to that arrived at by Gonzalez \& Laws (2000);
it is believed that the field star sample used by them contained too wide
a range of ages, evolutionary types, and temperatures to be accommodated by the
model they
adopted to explain the dependence on parameters. 

Li does not appear set to provide key insights into the formation and evolution
of planetary systems.

% \section{Acknowledgements}

\bsp

\label{lastpage}

\end{document}